\documentclass[12pt]{elsarticle}
\usepackage{refmerge}
\usepackage{epsfig}
\begin{document}
\date{\today}

\title{\bf{ \boldmath
STUDY OF THE PROCESS $e^+e^-\to 3(\pi^+\pi^-)$
IN THE C.M.ENERGY RANGE 1.5--2.0 GEV WITH THE CMD-3 DETECTOR
}}

\author[adr1]{R.R.Akhmetshin}
\author[adr1]{A.V.Anisenkov}
\author[adr1,adr2]{S.A.Anokhin}
\author[adr1,adr2]{V.M.Aulchenko}
\author[adr1]{V.S.Banzarov}
\author[adr1]{L.M.Barkov}
\author[adr1]{N.S.Bashtovoy}
\author[adr1,adr2]{D.E.Berkaev}
\author[adr1,adr2]{A.E.Bondar}
\author[adr1]{A.V.Bragin}
\author[adr1,adr2]{S.I.Eidelman}
\author[adr1,adr6]{D.A.Epifanov}
\author[adr1,adr3]{L.B.Epshteyn}
\author[adr1,adr2]{G.V.Fedotovich}
\author[adr1,adr2]{S.E.Gayazov}
\author[adr1,adr2]{A.A.Grebenuk}
\author[adr1,adr3]{D.N.Grigoriev}
\author[adr1]{E.N.Gromov}
\author[adr1]{F.V.Ignatov}
\author[adr1]{S.V.Karpov}
\author[adr1,adr2]{V.F.Kazanin}
\author[adr1,adr2]{B.I.Khazin}
\author[adr1,adr2]{I.A.Koop}
\author[adr1]{A.N.Kozyrev}
\author[adr1,adr2]{P.P.Krokovny}
\author[adr1,adr3]{A.E.Kuzmenko}
\author[adr1]{A.S.Kuzmin}
\author[adr1,adr2]{I.B.Logashenko}
\author[adr1]{A.P.Lysenko}
\author[adr1,adr2]{P.A.Lukin}
\author[adr1]{K.Yu.Mikhailov}
\author[adr1]{Yu.N.Pestov}
\author[adr1,adr2]{E.A.Perevedentsev}
\author[adr1]{S.A.Pirogov}
\author[adr1]{S.G.Pivovarov}
\author[adr1,adr2]{A.S.Popov}
\author[adr1]{Yu.S.Popov}
\author[adr1]{S.I.Redin}
\author[adr1]{Yu.A.Rogovsky}
\author[adr1]{A.L.Romanov}
\author[adr1]{A.A.Ruban}
\author[adr1]{N.M.Ryskulov}
\author[adr1,adr2]{A.E.Ryzhenenkov}
\author[adr1]{V.E.Shebalin}
\author[adr1,adr2]{D.N.Shemyakin}
\author[adr1,adr2]{B.A.Shwartz}
\author[adr1]{D.B.Shwartz}
\author[adr1,adr4]{A.L.Sibidanov}
\author[adr1]{P.Yu.Shatunov}
\author[adr1]{Yu.M.Shatunov}
\author[adr1]{\fbox{I.G.Snopkov}}
\author[adr1,adr2]{E.P.Solodov\fnref{tnot}}
\author[adr1]{V.M.Titov}
\author[adr1,adr2]{A.A.Talyshev}
\author[adr1]{A.I.Vorobiov}
\author[adr1]{Yu.V.Yudin}
\author[adr1,adr5]{A.S.Zaytsev}

\address[adr1]{Budker Institute of Nuclear Physics, SB RAS, 
Novosibirsk, 630090, Russia}
\address[adr2]{Novosibirsk State University, Novosibirsk, 630090, Russia}
\address[adr3]{Novosibirsk State Technical University, 
Novosibirsk, 630092, Russia}
\address[adr4]{University of Sydney, School of Physics, 
Falkiner High Energy Physics, NSW 2006, Sydney, Australia}
\address[adr5]{Brookhaven National Laboratory, P.O. Box 5000 Upton, 
NY 11973-5000, USA}
\address[adr6]{University of Tokyo, Department of Physics, 
7-3-1 Hongo Bunkyo-ku Tokyo, 113-0033, Japan}
\fntext[tnot]{Corresponding author:solodov@inp.nsk.su}


%
\vspace{0.7cm}
\begin{abstract}
\hspace*{\parindent}
The cross section of the process $e^+e^- \to 3(\pi^+\pi^-)$ has been measured
using 22 pb$^{-1}$ of integrated luminosity collected with the CMD-3
detector at the VEPP-2000  $e^+e^-$ collider 
in the c.m. energy range 1.5 -- 2.0 GeV.
The measured cross section exhibits a sharp drop near the $p\bar
p$ threshold. A first study of dynamics of six-pion production 
has been performed. 
\end{abstract}

\maketitle
\baselineskip=17pt
\section{ \boldmath Introduction}
\hspace*{\parindent}
Production of six pions in $e^+e^-$ annihilation
was studied at DM2~\cite{6pidm2} and with much larger  
effective integrated luminosity at BaBar~\cite{isr6pi},
using Initial-State Radiation (ISR) events. The DM2 experiment
observed a ``dip'' in the cross section at about 1.9 GeV, confirmed
later by the FOCUS Collaboration 
in the photoproduction~\cite{focus, focus1}
and by the BaBar  Collaboration, where this structure was also observed in the
$2(\pi^+\pi^-)\pi^0\pi^0$ final state~\cite{isr6pi}. 
The origin of the ``dip'' remains unclear, but the most popular
explanation suggests a presence of the under-threshold proton-antiproton
($p\bar p$) resonance. This hypothesis is supported by the fast increase of
the $p\bar p$ form factor to the threshold, recently confirmed by the
high-statistics BaBar study~\cite{isrppbar}, and discussed in many
theoretical papers (see, e.g., Ref.~\cite{ppbartheory}).
Even earlier, a narrow structure near the  proton-antiproton threshold 
has been also observed in the total cross section of $e^+e^-$ annihilation 
into hadrons in the FENICE experiment~\cite{fenice}.
  
The $e^+e^- \to 3(\pi^+\pi^-)$ cross section is also used  
in the calculation of 
the hadronic contribution to the muon anomalous magnetic 
moment~\cite{g-2}. The detailed study of the production dynamics can
further improve the accuracy of these calculations and help in 
explaining the cross section  anomaly.

In this paper we report the analysis of the data sample based on 
33 pb$^{-1}$ of integrated luminosity collected at the CMD-3 detector
in the 1.0-2.0 GeV center-of-mass energy range. We observe only a few
candidate events below 1.5 GeV. Since their number is consistent with
background, we present our results
for the 1.5-2.0 GeV center-of-mass energy range, corresponding
to 22  pb$^{-1}$ of integrated luminosity. 
These data were collected in three energy scans performed at 
the VEPP-2000 collider~\cite{vepp}.

The general purpose detector CMD-3 has been described in 
detail elsewhere~\cite{sndcmd3}. Its tracking system consists of a 
cylindrical drift chamber (DC)~\cite{dc} and double-layer multiwire 
proportional 
Z-chamber, both also used for a trigger, and both inside a thin 
(0.2~X$_0$) superconducting solenoid with a field of 1.3~T.
The liquid xenon (LXe) barrel calorimeter with 5.4~X$_0$ thickness has
fine electrode structure, providing good spatial resolution~\cite{lxe}, and
shares the cryostat vacuum volume with the superconducting solenoid.     
The barrel CsI crystal calorimeter with thickness 
of 8.1~X$_0$ is placed
outside  the LXe calorimeter,  and the end-cap BGO calorimeter with a 
thickness of 13.4~X$_0$ is placed inside the solenoid~\cite{cal}.
The luminosity is measured using events of Bhabha scattering 
at large angles~\cite{lum}. 
\section{Selection of $e^+e^-\to 3(\pi^+\pi^-)$ events}
\label{select}
\hspace*{\parindent}

Candidates for the process under study 
are required to have 
five and more charged-particle tracks with 
the following ``good'' track definition:
\begin{itemize}
\item{
A track contains more than five hits in the DC.
}
\item{
A track momentum is larger than 40 MeV/c.
}
\item{
A minimum distance from a track to the beam axis in the
transverse  plane is less than 0.5 cm.
}
\item{
A minimum distance from a track to the center of the interaction region along
the beam axis Z  is less than 10 cm.
}
\item{
A track has a polar angle large enough to cross half of the DC radius.
}
\end{itemize}

The number of events with
seven or more selected  tracks is found to be less
than 1\%. 
Reconstructed momenta 
and angles of the tracks for six-track and five-track events were used
for further selection.

\begin{figure}[tbh]
\begin{center}
\vspace{-0.7cm}
\includegraphics[width=0.8\textwidth]{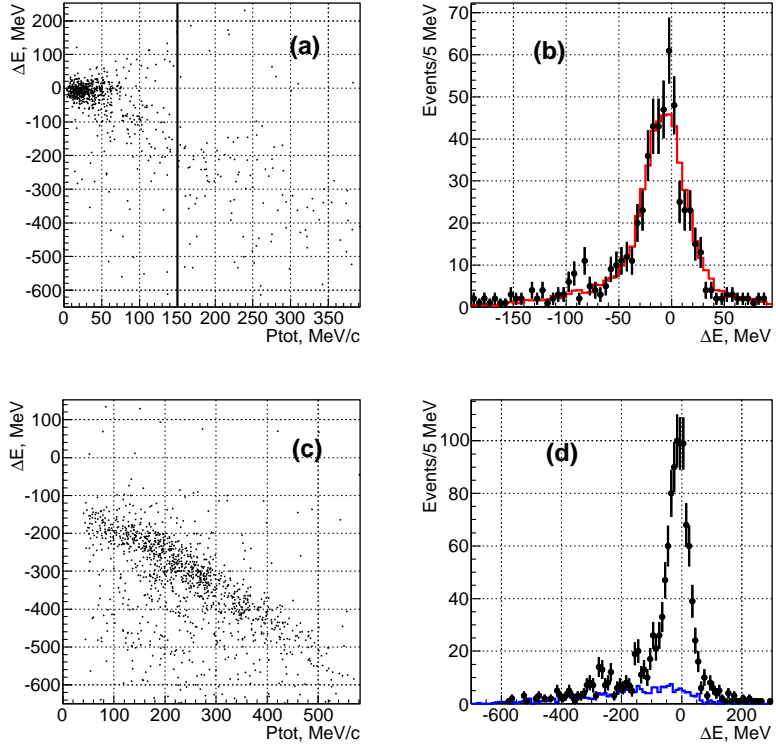}
\vspace{-0.3cm}
\caption
{
(a) Scatter plot of the difference ($\Delta E$) between the total energy  and
  c.m. energy
  versus total momentum for six-track events. The line shows 
the boundary of the applied selection; 
(b) Projection plot of (a) after selection. The histogram shows 
the normalised MC-simulated distribution;  
(c) Scatter plot of difference ($\Delta E$) of total energy 
and c.m. energy
  versus total momentum for five-track events; 
(d) Difference between the total energy of
  five-tracks plus missing track energy  
  and c.m. energy (points). The histogram shows the
  distribution for the MC simulated background events (see text).    
}
\label{energy}
\end{center}
\end{figure}
For six- or five-track candidates we calculate the total energy and total
momentum assuming all tracks to be pions:
$$
\rm Etot = \sum_{i=1}^{5,6}\sqrt{p_{i}^2+m_{\pi}^2}~,  ~~~~~Ptot =
\large |\sum_{i=1}^{5,6}\bar p_{i}\large |.
$$

Figure~\ref{energy}(a) shows a scatter plot of the difference between 
the total energy and c.m. energy $\Delta$E=Etot--Ec.m. versus total
momentum for six-track candidates. 
The histograms combine events from three highest energy points.
A clear signal of six-pion events is seen as a cluster of dots 
near zero.
Events with a radiative photon have non-zero total momentum and  total
energy which is  always smaller than the nominal one.    
A momentum of any pion incorrectly reconstructed due to interaction with
detector material or DC resolution leads to momentum-energy
correlated ``tails'' in both directions.   

We select events with total momentum less than 150 MeV/c and
show the difference $\Delta$E in Fig.~\ref{energy}(b). 
The experimental points are in good agreement with the
corresponding Monte Carlo (MC) simulated distribution shown by the histogram.
We require -200$<\Delta$E$<$100 MeV 
to determine the number of six-pion events. 
Six-track events have
practically no background: we estimate it from MC simulation 
of the major background  
processes $2(\pi^+\pi^-\pi^0)$ and $2(\pi^+\pi^-)\pi^0$
(one  of the photon from
the $\pi^0$ decay converts to a $e^+ e^-$ pair at the vacuum pipe),
and found a contribution of
less than 1\%. We use this value as an estimate of the 
corresponding systematic uncertainty. 

To determine the number of six-pion events with one missing track, a sample
with five selected tracks is used. A track can be lost if it 
flies at  small polar angles outside the 
efficient DC region, decays in flight, due to incorrect
reconstruction, nuclear interactions or by overlapping with 
another track. 
Figure~\ref{energy}(c) shows a scatter plot of the difference 
$\Delta$E between the total
energy and c.m. energy versus total
momentum for five-track events. 
Six-pion candidates in the five-track sample have energy
deficit correlated with the total momentum.
This sample has some admixture of background events from
multihadron processes mentioned above with photons from the $\pi^0$ decays.
We apply an additional requirement on the
``neutral'' (not associated with charged tracks)  energy 
$E_{neutral}$ in the calorimeter 
to be less than 300 MeV. This requirement 
reduces the background by a factor of two and
removes less than 2\% of
signal events estimated using MC simulation. 
 
The direction and  momentum of a missing pion can be calculated
assuming a  six-pion final state.
We add energy of a missing pion to the energy of five detected
pions and show the difference $\Delta$E 
in Figure~\ref{energy}(d) by points.
A corresponding background distribution from the MC simulation of the
$2(\pi^+\pi^-\pi^0)$ and $2(\pi^+\pi^-)\pi^0$ events
is shown in Figure~\ref{energy}(d) by the histogram: 
background events contribute
less or about 10\% to the signal region after applying a 
requirement $E_{neutral}<$300 MeV.   

To obtain the number of six-pion events from the five-track sample,
we fit the distribution shown in Fig~\ref{energy}(d) with a sum of
functions describing a signal peak and background.
The signal line shape is taken from the MC simulation 
of the six-pion process and is well described by a sum of two 
Gaussian distributions.
The photon emission by initial electrons and positrons is taken into
account in
the MC simulation and gives a small asymmetry observed in the 
distributions of Figs.~\ref{energy} (b,d).
 We describe this asymmetry by an admixture of a third Gaussian function. 
All parameter ratios of the signal function are fixed except for the number of
events and main Gaussian resolution.
The third-order polynomial is used to describe the background distribution. 

To estimate a systematic uncertainty of the background subtraction procedure,
we compare the MC simulated background distribution with the
experimental events with an $E_{neutral}>300$ MeV requirement, and 
found reasonable agreement
with the histogram shown in Fig~\ref{energy}(d). 
A variation of the polynomial fit parameters for the experimental and MC
simulated background distributions leads to about 3\% uncertainty on the
number of signal events.

We found 2887 six-track events and 5069 five-track events
corresponding to the process $e^+e^-\to 3(\pi^+\pi^-)$. 
The numbers of six- ($N_{6\pi}$) and five-track ($N_{5\pi}$) events
determined at each energy point are listed in Table~\ref{table}.
\section{First study of the production dynamics}
\label{dynamics}
\hspace*{\parindent}
To obtain a detection efficiency, we simulate six-pion production 
in a primary generator, pass simulated events through the 
CMD-3 detector using the GEANT4~\cite{geant4} package, and reconstruct them
with the same reconstruction software as experimental data. 
In our experiment,
the acceptance of the DC for the charged tracks is not 100\%, and
the detection efficiency depends on the production dynamics of six pions.
The dynamics of the process $e^+e^-\to 3(\pi^+\pi^-)$   
was not studied previously in detail. The BaBar Collaboration~\cite{isr6pi}
reported the observation of only one $\rho(770)$ from all $\pi^+\pi^-$
invariant mass combinations and no structures in any other (three-, four-pion)
invariant mass combinations.

We investigate a few
 production mechanisms, and
compare simulated angular and invariant mass distributions with those
in data. 
All studied distributions strongly
contradict to a phase space model, which assumes all pions to be completely
independent. We exclude the phase space model from further consideration.
In this paper we illustrate our study with three models, all with one 
$\rho(770)$
per event. To conserve the initial state quantum numbers, 
six pions must have $J^{PC}=1^{- -}$.

\begin{center}
\begin{figure}[tbh]
\vspace{-0.2cm}
\includegraphics[width=1.0\textwidth]{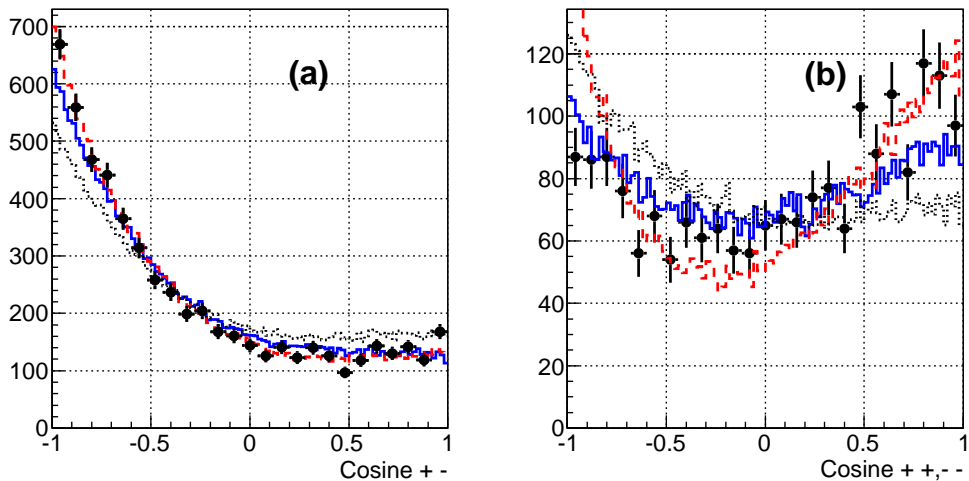}
\caption
{
Cosines of the relative angle of two pions with opposite-sign charge
  (a), and of two pions with same-sign charge (b) for experimental events
(dots) and MC simulation for $\rho(1420)\pi^+\pi^-\to a_1(1260)\pi\pi^+\pi^-$ 
(dotted histogram),  $\rho(770)f_0(1370)$ (solid histogram) and
$\rho(770)f_2(1270)$ (dashed histogram).
}
\label{cosine}
\end{figure}
\end{center}
In the model \#1 we use the following decay
chain: $e^+e^-\to \rho(1420)(\pi^+\pi^-)_{S-wave} \to
a_1(1260)^{\pm}\pi^{\mp}\pi^+\pi^- \to \rho(770)^0 2(\pi^+\pi^-) \to   
3(\pi^+\pi^-)$. This model uses dominant decays
$\rho(1420)^0\to a_1(1260)^{\pm}\pi^{\mp}$
and $a_1(1260)^{\pm}\to\rho(770)^0\pi^{\pm}$~\cite{cmd2_4pi}, and 
naturally includes 
the $a_1(1260)^{\pm}\to\rho(770)^{\pm}\pi^0$ decay to describe 
the $e^+e^-\to 2(\pi^+\pi^-\pi^0)$ process with one charged 
$\rho(770)$~\cite{isr6pi}. We use PDG values~\cite{pdg} for the
resonance parameters and the model allows to introduce a form factor 
in each decay vertex.

Another studied model (\#2) was simpler: it includes
the production of one $\rho(770)^0$  
and four pions  in S-wave. 
We try two options: the four pions are
distributed according to the phase space or forming a scalar
resonances $f_0(1370)$ or $f_0(1500)$. 

And finally, the model (\#3) assumes $e^+e^-\to\rho(770)f_2(1270)$ with
a tensor $f_2$ resonance in the four-pion final state.

MC simulation should reproduce experimental angular distributions of the pions
to obtain correct detection efficiency. 
Figure~\ref{cosine} shows (by points) the
cosines of open angles between pions for opposite-sign (a)  and
same-sign (b) pion pairs for data.

We compare distributions of Fig.~\ref{cosine} with the 
MC simulated  distributions 
for the model \#1 (dotted histogram), 
model \#2 (solid histogram) and model \#3 (dashed histrogram), 
and the best agreement was found with the model \#2.

Note, that variation of the resonance parameters in the models does not
significantly affect these angular distributions. For example, model
\#2 with production of one $\rho(770)^0$ exhibits the same angular
distributions both in the case, when 
the remaining four pions are distributed according 
to phase space or form a scalar
resonance ($f_0(1370)$ or $f_0(1500)$). 

\begin{center}
\begin{figure}[tbh]
\vspace{-0.2cm}
\includegraphics[width=1.0\textwidth]{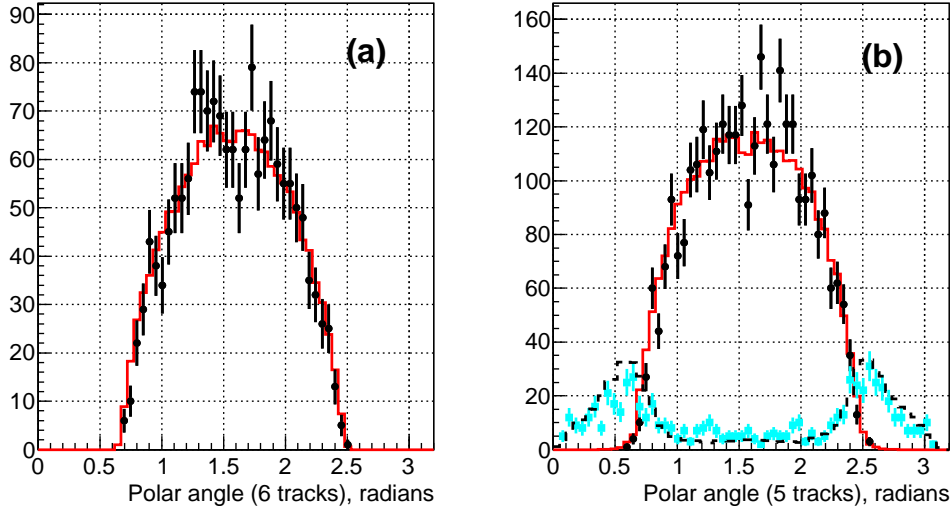}
\caption
{
(a) Polar angle distribution for six-pion events with six detected
  tracks for data (points) and MC simulation (histogram);
(b) Polar angle distribution for six-pion events with five detected
  tracks for data (circles) and MC simulation (solid histogram). 
The polar angle distribution for a missing track is shown by squares (data)
and the dashed histogram (MC simulation).
}
\label{angles}
\end{figure}
\end{center}
\begin{center}
\begin{figure}[p]
\vspace{-0.2cm}
\includegraphics[width=0.95\textwidth]{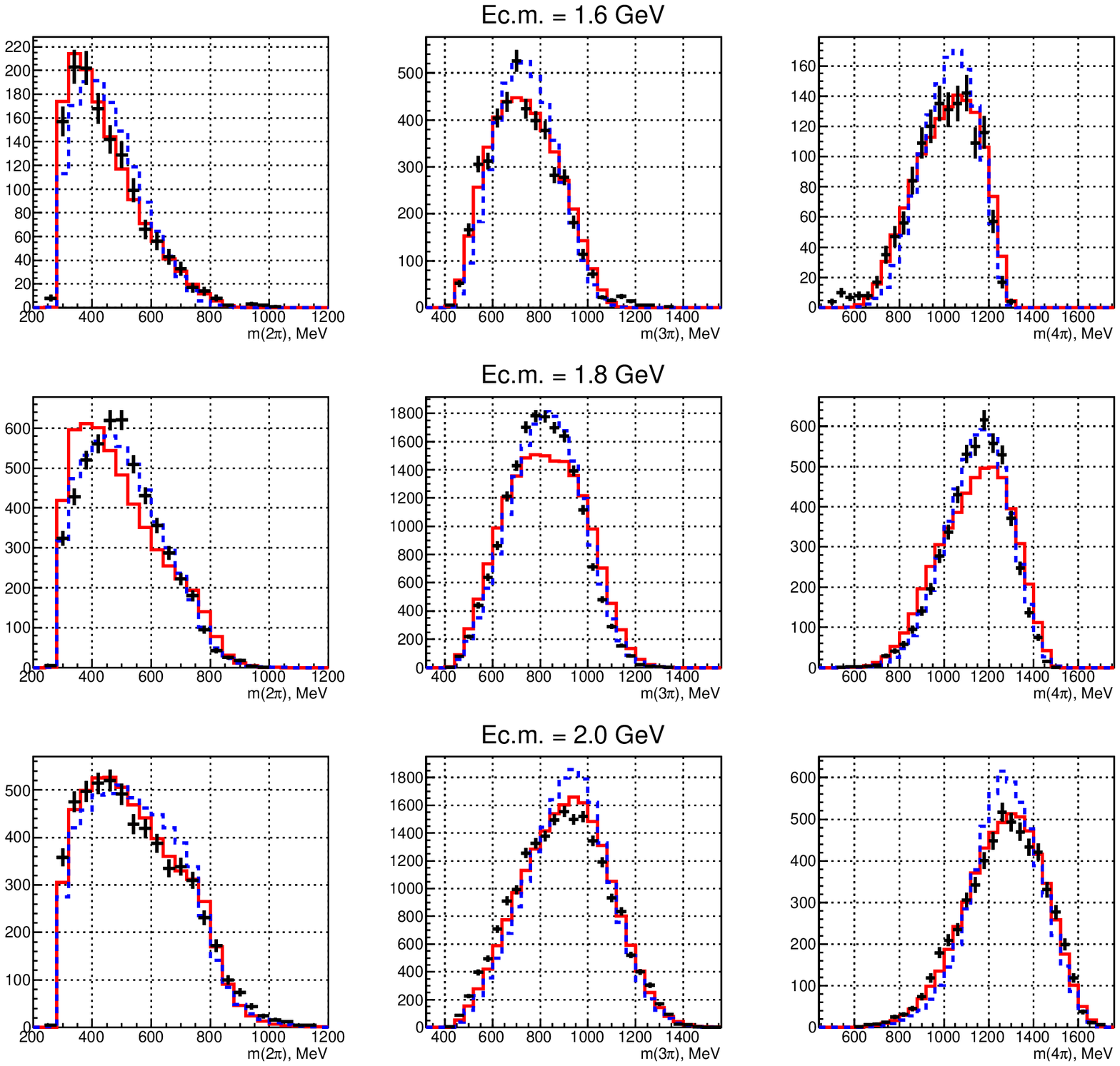}
\caption
{Experimental
invariant mass distributions (from left to right) for two, three and
four pions for (top to bottom) 1600, 1800 and 2000 MeV c.m. energies in
comparison with  simulation of one $\rho(770)^0$  
with the remaining four pions in S-wave,
and distributed according to the phase space (solid histogram) or form a scalar
resonance $f_0(1370)$ (dashed histogram).  
}
\label{masses}
\end{figure}
\end{center}

Figure~\ref{angles}(a) presents the polar angle ($\theta_{\pi}$)
distribution for six-pion events with all detected
tracks. The requirement for a track 
to cross half of the DC radius effectively  determines a cut on this 
parameter. The result of the MC simulation in model \#2, presented by
the histogram,  
well describes the observed distribution.
Figure~\ref{angles}(b) presents the polar angle distribution for
five detected tracks (circles for data, the solid
histogram for the MC simulation) after background subtraction.
 The polar angle distribution for the missing track
is shown by squares (data) and the dashed histogram (MC).
With our ``effective'' DC acceptance we have almost 
two times more six-pion events with one missing track than events
with all tracks detected.

We calculate invariant masses for the combinations of 
two, four (total charge zero), and  
three (total charge $\pm 1$) pions for the different
c.m. energies and show them in Fig.~\ref{masses}. We compare the
obtained distributions with model \#2 ($\rho 4\pi$),
and observe good agreement with experiment at c.m. energies 1600 MeV
and 2000 MeV, if four pions are distributed according to phase space
(solid histogram). 
But at the c.m. energy of 1800 MeV the experimental data are better
decsribed by the same model with four pions forming $f_0(1370)$.
Note that invariant mass distributions for models \#1 and \#3 do not
describe data in any mass interval, but some admixture of these channels
cannot be excluded.

From the study of the mass distributions in 
Fig.~\ref{masses} we conclude that production dynamics of six charged
pions changes in the relatively narrow energy region (1700-1900
MeV). This phenomenon demands a further investigation.
\section{Detection efficiency}
\label{efficiency}
\hspace*{\parindent}
We calculate the detection efficiency from the MC simulated events
as a ratio of events after selections described in Sec.~\ref{select} 
to the total number of generated events. With the limited DC
acceptance, incorrect simulation of the pion angular distribution leads to a
systematic error in the efficiency calculation and thus in the
cross section measurement.
   
\begin{center}
\begin{figure}[tbh]
\vspace{-0.2cm}
\includegraphics[width=1.0\textwidth]{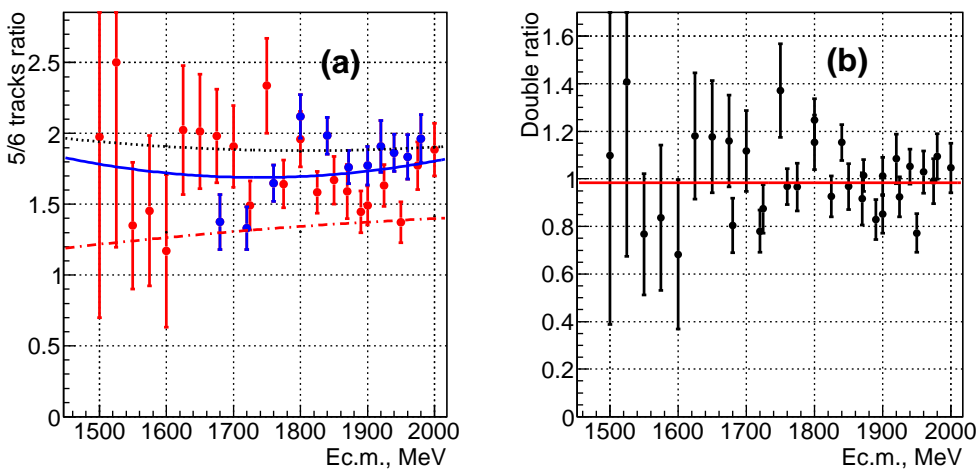}
\caption
{
(a) Ratio of events with five and six detected tracks for data (points
  with errors) and MC simulation for model \#1 (dotted line), model
  \#2 (solid line) and model \#3 (dashed line);
(b) Double ratio $R_{56}^{data}$/$R_{56}^{MC\#2}$ versus energy.
The line shows a fit with a constant.
}
\label{ratio}
\end{figure}
\end{center}

In the five-track sample, about 15-17\% of events
have a missing track due to the DC reconstruction inefficiency, well
reproduced by the MC 
simulation. The remaining events migrate from the six- to
the five-track sample due to the limited DC acceptance (see Fig.~\ref{angles}). 
It makes the ratio $R_{56} = N_{5\pi}/N_{6\pi}$ very sensitive to the 
pion angular distribution, and we study it to validate the model used
for the efficiency calculation.

Figure~\ref{ratio} (a) shows the $R_{56}$ ratio versus energy for 
data (points with errors) and for three models, discussed in 
Sec.~\ref{dynamics}. 
The experimental average
value  $R_{56}^{data}$ = 1.74$\pm$0.03 is in good agreement with
$R_{56}^{MC\#2}$ = 1.76 for the model \#2 (solid line), but
inconsistent with model \#1 ($R_{56}^{MC\#1}$ = 1.92, dotted line) and model \#3
($R_{56}^{MC\#3}$ = 1.30, dashed line).   
A ``naive'' phase space model 
for the six-pion production (all tracks uncorrelated) gives
$R_{56}^{MC}$ = 2.1. 

To estimate a model-dependent systematic error, we compare 
the experimental number of six- and five-track events after
normalisation to the MC simulated acceptance. 
We calculate a 
double ratio $R_{56}^{data}/R_{56}^{MC\#2}$
for each energy point for the model \#2,
and show it in Fig.~\ref{ratio} (b). 
The average value 0.984$\pm$0.018 ($\chi^2/n.d.f$=56/35)
is in good agreement  with the 
prediction of model \#2 in the studied energy interval, so that a
maximum systematic deviation from unity does not exceed 3.4\%.
However, a relatively large $\chi^2$ value can be an indication of the
additional systematic uncertainty, and 
we conservatively take 4\% as an estimate of a
systematic error on the detection efficiency using $\sqrt{\chi^2/n.d.f}$
as a scale factor.
\begin{center}
\begin{figure}[tbh]
\vspace{-0.2cm}
\includegraphics[width=1.0\textwidth]{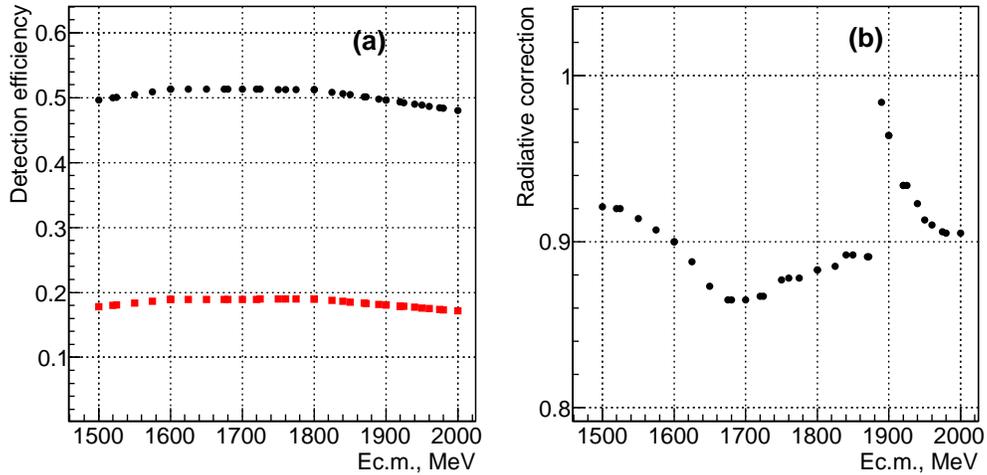}
\caption
{
(a) Efficiency calculated from the MC simulation for six-track events
  (squares) and for a sum of five- and six-track events (circles);
(b) Radiative correction.
}
\label{eff}
\end{figure}
\end{center}

The detection 
efficiency thus obtained with model \#2 is shown in Fig.~\ref{eff}(a) 
for events
with six detected tracks (squares) and for a sum of five- and
six-track events (circles), icreasing efficiency by factor 2.5.
Note that if a sum of six- and five-track events ($N_{6\pi} +
N_{5\pi}$)  is taken for the detection 
efficiency calculation, 
the data-MC inconsistencies in the description of the DC inefficiency 
and (partly) in the model-dependent angular distributions are
significantly reduced.

\section{Cross Section Calculation}    
\hspace*{\parindent}
At each energy the cross section is calculated as 
$$
\sigma = \frac{N_{6 \rm tr}+N_{5 \rm tr}}{L\cdot\epsilon\cdot(1+\delta)},
$$ 
where 
$L$ is the integrated luminosity for this energy point, $\epsilon$ is 
the detection efficiency (Fig.~\ref{eff}(a)), 
and $(1+\delta)$ is the radiative correction calculated 
according to~\cite{kur_fad} and shown in Fig.~\ref{eff} (b).
The energy dependence of the radiative correction reflects a sharp dip 
in the cross section. To calculate the correction we use BaBar
data~\cite{isr6pi} as a first 
approximation and then use our cross section data for iterations.  

The integrated luminosity, the number of six and five-track events,
detection efficiency, radiative correction  and obtained cross section
for each energy point are listed in Table~\ref{table}. 

\begin{center}
\begin{figure}[tbh]
\vspace{-0.2cm}
\includegraphics[width=1.0\textwidth]{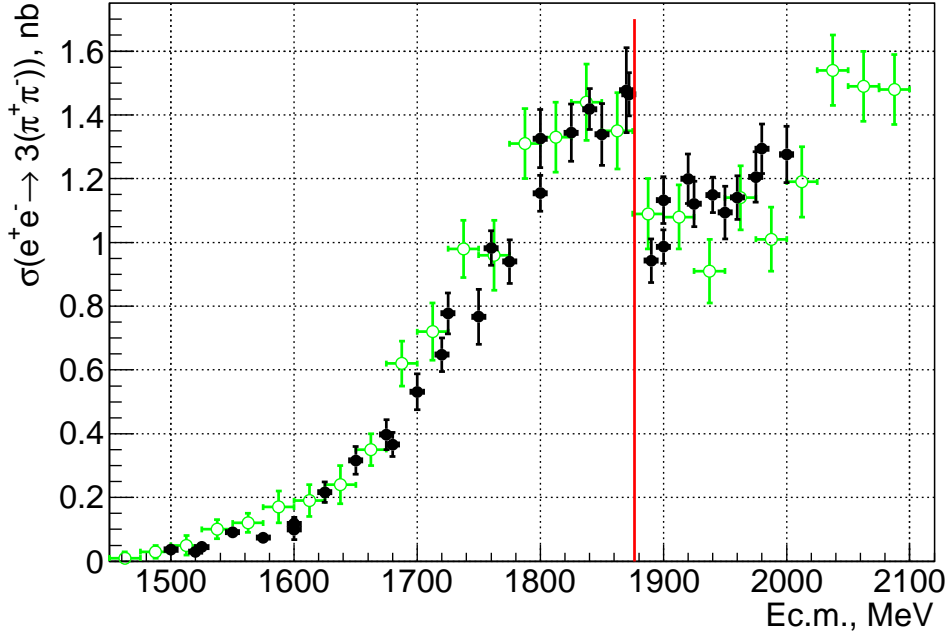}
\caption
{
The $e^+e^-\to 3(\pi^+\pi^-)$ cross section measured with the CMD-3
detector at VEPP-2000 (dots). The results of the BaBar
measurement~\cite{isr6pi} are shown by open circles. The line shows the
$p\bar p$ threshold.
}
\label{cross}
\end{figure}
\end{center}

\begin{table}[tb]

\caption{Luminosity, Number of events, Detection efficiency,
  Rad. correction and Cross section for each c.m. energy
  point. Horizontal lines separate three energy scans.}
\label{table}
\smallskip
\begin{center}
\renewcommand{\arraystretch}{0.95}
\begin{tabular}{ccccccc}
\hline
\hline
{\bf Ec.m. , MeV}&{\bf L, nb$^{-1}$}&{\bf\boldmath $N_{6\pi}$}&{\bf\boldmath $N_{5\pi}$}
&{\bf\boldmath $\epsilon_{MC}$ } &   {\bf\boldmath $1+\delta$ }&{\bf\boldmath $\sigma$, nb} \\
\hline
2000 & 474.7 & 88 & 166.0$\pm$14.8 & 0.480 & 0.905 & 1.28$\pm$0.09\\
1975 & 516.5 & 95 & 168.4$\pm$14.3 & 0.484 & 0.906 & 1.20$\pm$0.08\\
1950 & 458.8 & 91 & 124.8$\pm$13.2 & 0.488 & 0.913 & 1.09$\pm$0.08\\
1925 & 582.2 & 110 & 179.4$\pm$15.0 & 0.492 & 0.934 & 1.12$\pm$0.07\\
1900 & 495.6 & 104 & 155.1$\pm$13.5 & 0.496 & 0.964 & 1.13$\pm$0.07\\
1850 & 431.8 & 94 & 156.9$\pm$15.3 & 0.504 & 0.892 & 1.34$\pm$0.10\\
1800 & 440.1 & 86 & 168.6$\pm$15.0 & 0.513 & 0.883 & 1.33$\pm$0.09\\
1750 & 541.8 & 54 & 126.2$\pm$18.9 & 0.513 & 0.877 & 0.77$\pm$0.09\\
1700 & 486.1 & 38 & 72.5$\pm$10.0 & 0.513 & 0.865 & 0.53$\pm$0.06\\
1650 & 463.3 & 21 & 42.3$\pm$7.5 & 0.513 & 0.873 & 0.32$\pm$0.04\\
1600 & 441.9 & 9 & 10.5$\pm$5.5 & 0.513 & 0.900 & 0.099$\pm$0.032\\
1550 & 521.1 & 9 & 12.1$\pm$4.0 & 0.505 & 0.914 & 0.091$\pm$0.013\\
1500 & 554.6 & 3 & 5.9$\pm$4.1 & 0.497 & 0.921 & 0.037$\pm$0.018\\
\hline
1890 & 521.5 & 95 & 137.4$\pm$13.7 & 0.498 & 0.984 & 0.94$\pm$0.07\\
1870 & 663.4 & 163 & 259.1$\pm$35.9 & 0.501 & 0.891 & 1.48$\pm$0.13\\
1825 & 500.8 & 113 & 179.1$\pm$16.5 & 0.509 & 0.885 & 1.34$\pm$0.09\\
1775 & 550.7 & 85 & 139.7$\pm$13.5 & 0.513 & 0.878 & 0.94$\pm$0.07\\
1725 & 523.0 & 70 & 104.6$\pm$11.7 & 0.513 & 0.867 & 0.78$\pm$0.06\\
1675 & 561.4 & 32 & 63.4$\pm$9.8 & 0.513 & 0.865 & 0.40$\pm$0.05\\
1625 & 508.5 & 16 & 32.4$\pm$6.1 & 0.513 & 0.888 & 0.22$\pm$0.03\\
1575 & 522.2 & 7 & 10.2$\pm$3.5 & 0.509 & 0.907 & 0.074$\pm$0.011\\
1525 & 530.9 & 3 & 7.5$\pm$3.3 & 0.501 & 0.920 & 0.045$\pm$0.016\\
\hline
1980 & 602.2 & 111 & 217.9$\pm$16.5 & 0.484 & 0.905 & 1.29$\pm$0.08\\
1960 & 680.1 & 117 & 214.6$\pm$16.7 & 0.487 & 0.910 & 1.14$\pm$0.07\\
1940 & 988.7 & 173 & 322.4$\pm$20.2 & 0.490 & 0.923 & 1.15$\pm$0.06\\
1920 & 491.5 & 90 & 171.8$\pm$14.0 & 0.493 & 0.934 & 1.20$\pm$0.08\\
1900 & 883.3 & 145 & 257.1$\pm$17.7 & 0.496 & 0.964 & 0.99$\pm$0.05\\
1872 & 845.6 & 193 & 340.0$\pm$20.2 & 0.501 & 0.891 & 1.46$\pm$0.07\\
1840 & 952.1 & 197 & 390.7$\pm$22.4 & 0.506 & 0.892 & 1.42$\pm$0.06\\
1800 & 972.1 & 157 & 332.6$\pm$20.6 & 0.513 & 0.883 & 1.15$\pm$0.06\\
1760 & 950.4 & 153 & 252.2$\pm$18.7 & 0.513 & 0.878 & 0.98$\pm$0.05\\
1720 & 797.4 & 95 & 126.5$\pm$15.3 & 0.513 & 0.867 & 0.65$\pm$0.05\\
1680 & 879.2 & 58 & 79.7$\pm$12.0 & 0.513 & 0.865 & 0.37$\pm$0.04\\
1600 & 812.7 & 10 & 32.4$\pm$6.5 & 0.513 & 0.900 & 0.117$\pm$0.020\\
1520 & 825.3 & 2 & 8.9$\pm$3.6 & 0.500 & 0.920 & 0.030$\pm$0.011\\
\hline
\end{tabular}
\end{center}
\end{table}

\section{Systematic errors}
\hspace*{\parindent}
The following sources of systematic uncertainties are considered.

\begin{itemize}

\item {
The model dependence of the acceptance is determined using the angular
distributions, which are specific for each particular model. 
As shown in Sec.~\ref{efficiency}, a model with one $\rho(770)$ and
remaining pions in S-wave (phase space or $f_0(1370)$) gives good
overall agreement with the observed angular distributions. 
Using the ratio of six- and five-track events we estimate a systematic
uncertainty on the detection efficiency as 4\%.
}  
\item{
Since only one charged track is 
sufficient for a trigger (99-98\% efficiency), we assume 
that for the multi-track events, considered in this analysis, the 
trigger inefficiency gives a negligible contribution to the systematic error. 
}
\item{
A systematic error due to the selection criteria
is studied by 
varying the cuts described previously and doesn't exceed 3\%. 
}
\item{
The uncertainty on the determination of the integrated luminosity 
comes from the selection criteria of Bhabha events, radiative
corrections and calibrations of DC and CsI and does not exceed 
2\%~\cite{lum}.
}
\item{
The admixture of the background events not subtracted from 
the six-track sample is estimated as 1\%. 
}
\item{
The accuracy of background subtraction for five-track events is studied
by the variation of functions used for a background description in 
Fig.~\ref{energy}(d) and is estimated as 3\%. 
}
\item{
A possible uncertainty on the beam energy is studied using the momentum
distribution of Bhabha events and total energy of four-pion
events. The uncertainty at the level of $5\cdot10^{-3}$ is not excluded and
because of the cross section variation it can
result in a 1\% change of the cross section.}
\item{
A radiative correction uncertainty is estimated as about 
1\% mainly due to the uncertainty on the maximum allowed energy of the 
emitted photon, as well as from the uncertainty on the cross section.
}
\end{itemize}

The above systematic uncertainties summed in quadrature give an overall
systematic error of about 6\%. 

The obtained cross section is in overall agreement with the results 
of the most precise
measurement performed by the BaBar Collaboration~\cite{isr6pi} shown
in Fig.~\ref{cross} by open circles.

\section*{ \boldmath Conclusion}
\hspace*{\parindent}
The total cross section of the process $e^+e^-\to 3(\pi^+\pi^-)$ 
has been measured using 22 pb$^{-1}$ of integrated luminosity 
collected by the CMD-3 detector at the VEPP-2000 $e^+e^-$ collider
in the 1.5-2.0 GeV c.m. energy range. 
The five- and six-track events are used to estimate 
the model-dependent uncertainty in the acceptance calculation.
From our study  we can 
conclude that the observed production mechanism can be described by
the production of one $\rho(770)$ with four remaining pions in
S-wave and distributed according to  phase space. 
We also observe that the production dynamics changes 
in the 1700-1900 MeV c.m.energy range and demands further investigation.
A detailed analysis of the production dynamics 
will be performed in the combined analysis of the processes
$e^+e^-\to 3(\pi^+\pi^-)$ and $e^+e^-\to 2(\pi^+\pi^-)2\pi^0$. 

The measured cross section is in good agreement with all
previous experiments in the energy range studied, and exhibits a 
sharp dip near the $p\bar p$ threshold.

\subsection*{Acknowledgements}
\hspace*{\parindent}
The authors are grateful to A.I.~Milstein and Z.K.~Silagadze
for their help with a theoretical interpretation and development of
the models. 
We thank the VEPP-2000 team for excellent machine operation. 

This work is supported in part by the Russian Education and Science
Ministry, by FEDERAL TARGET PROGRAM "Scientific
  and scientific-pedagogical personnel of innovative Russia in
  2009-2013", by agreement 14.B37.21.07777,   by the Russian Fund for
  Basic Research grants   
RFBR 10-02-00695-a,
RFBR 10-02-00253-a, 
RFBR 11-02-00328-a, 
RFBR 11-02-00112-a, 
RFBR 12-02-31501-a,  
RFBR 12-02-31499-a,
RFBR 12-02-31498-a, 
and RFBR 12-02-01032-a.

\end{document}